\documentclass[aps,prl,twocolumn]{revtex4}
\usepackage{amssymb}
\usepackage{amsfonts}
\usepackage{amsmath}
\usepackage{graphicx}

\begin{document}

\title{Absence of Andreev Bound States in Noncentrosymmetric Superconductor PbTaSe$_2$ under Hydrostatic Pressures}

\author{Yu-qing Zhao$^1$, Zhi-fan Wu$^1$, Hai-yan Zuo$^1$, Wei-ming Lao, Yao He$^1$, Hai Wang$^1$ Ling-xiao Zhao$^2$, Ying-hui Sun$^2$, Huai-xin Yang$^2$, Geng-fu Chen$^{2,\ddag}$, Cong Ren$^{1,3}$}

\address{$^1$ School of Physics and Astronomy, Yunnan University, Kunming {\rm 650500}, China}
\address{$^2$ Beijing National Laboratory for Condensed Matter Physics, Institute of Physics, Chinese Academy of Science, Beijing {\rm 100190}, China}
\address{$^3$ Yunnan Key Laboratory for Electromagnetic Materials and Devices, Yunnan University, Kunming {\rm 650500}, China}

\begin{abstract}
Noncentrosymmetric superconductor PbTaSe$_2$, hosting bulk nodal-line fermions (Phys. Rev. B. 89, 020505) and spin-helical surface states (Nature Communication 7, 10556), represents a prime candidate for realizing topological superconductivity and Majorana bound states (MBS).  However, the definitive experimental signature of MBS in this system has thus far remained elusive.  Here we provide a comprehensive investigation of its superconducting properties under hydrostatic pressure. Combining Andreev reflection spectroscopy and temperature-dependent resistance measurements, we identify a separated surface-like superconductivity from the bulk one at a critical pressure $P_c$. The superconducting surface state demonstrate an $s$-wave pairing state with a strong coupling strength. Under magnetic fields, the absence of zero-bias conductance peak in the pressurized point-contact Andreev reflection spectrum. Our findings imposes a constraint on the theoretical proposals for realizing Majorana bound states in noncentrosymmetric superconductors.
\end{abstract}

\maketitle

\emph{Introduction:} Non-centrosymmetric superconductors (NCSs) exhibit unique physical phenomena due to the lack of spatial inversion center in crystal lattice. This symmetry-breaking leads to asymmetric spin-orbit coupling (ASOC), which causes the splitting of the Fermi surface and results in the mixing of even-parity/spin-singlet and odd-parity/spin-triplet pairing states. This mixing can give rise to novel topological phenomena, making NCSs promising candidates for topological superconductors (TSCs)\cite{Zhang2011,Kane2010}. Thus, even if the NCSs bulk pairing remains spin-singlet, a topologically protected surface state with spin-triplet pairing can emerge, facilitating the realization of TSCs. Importantly, the interplay between ASOC and superconductivity can be tuned, potentially enhancing the robustness of the surface states. NCSs are expected to display a range of unconventional topological properties, such as helical topological surface states (TSS) and topologically protected gap nodes. These features enhance NCSs potential as platforms for supporting Majorana fermions \cite{Beenakker2013, Beenakker2015, Vishwanath2011, Franz2015}, which are fundamental to topological quantum computing due to their non-Abelian statistics. As a result, NCS materials have garnered significant attention and have become the focus of extensive research.

\begin{figure}
\centering
\includegraphics[scale=0.38]{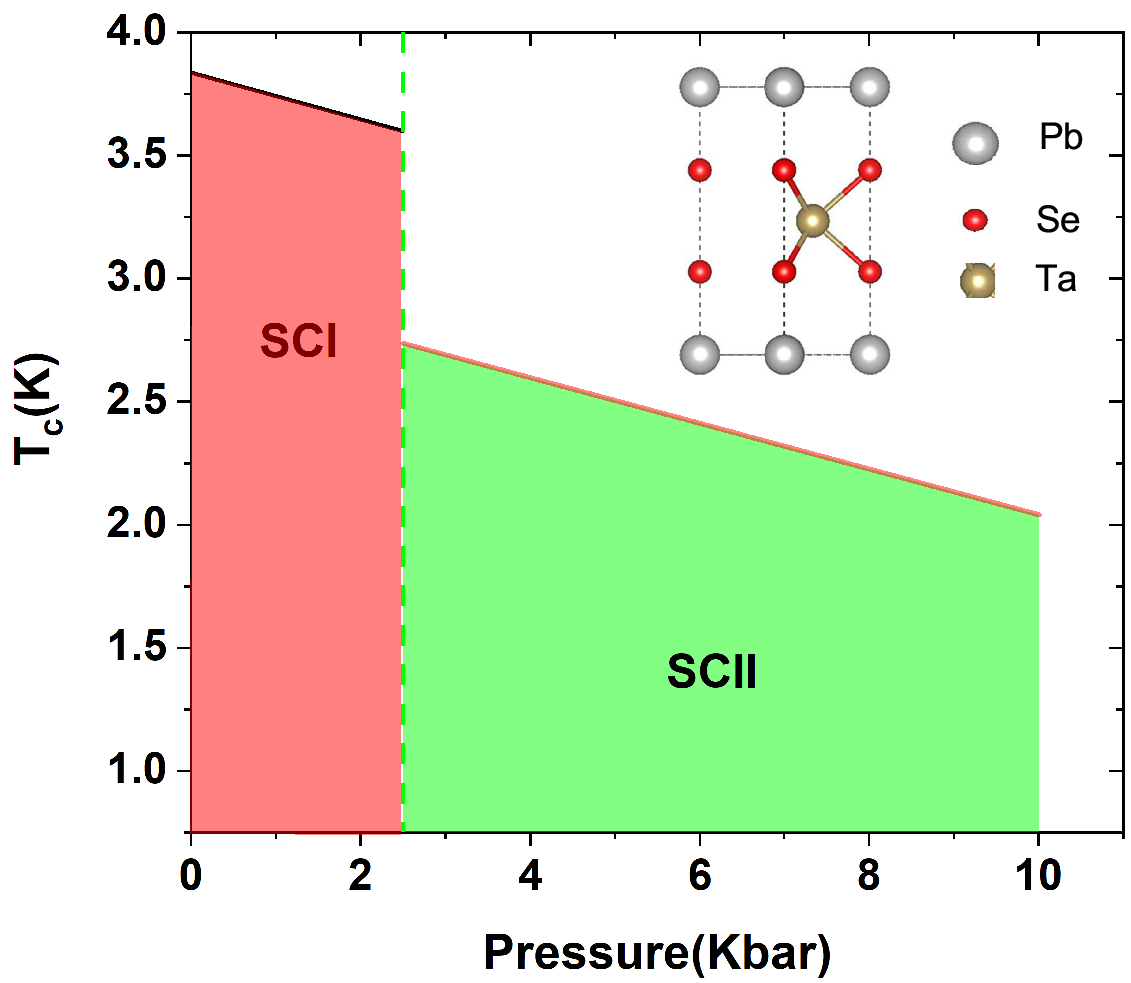}
\caption{The schematic of $T_c-P$ phase diagram for PbTaSe$_2$. Two regions, SCI and SCII are shown here based on Refs. \cite{Canfield,XuxiaoF}. Insert is the crystal structure of PbTaSe$_2$. }
\end{figure}

Chemically stoichiometric PbTaSe$_2$ is a typically non-centrosymmetric superconductor with $T_c=3.8$ K \cite{Cava2014}.  In its crystal structure, the Pb atoms intercalated TaSe$_2$ in which triangle lattices of Pb atoms layers are sandwiched between the hexagonal TaSe$_2$ layers. This alternately stacked structure is noncentrosymmetric, as schematically shown inset of Figure 1.  Thus, PbTaSe$_2$ has garnered significant attention as a strong candidate for topological superconductivity (TSC) \cite{Cava2014,Zheng2017, Yuan2016, Shuang2016, Luke2017, Li2016, Hanasaki2020, Jia2019, Iwasa2020, Tamega2020}. This material undergoes a superconducting transition below 3.8 K and crystallizes in the P6$\bar{m}2$ space group. Its layered structure consists of alternating TaSe$_2$ and Pb layers, where the incorporation of Ta atoms breaks inversion symmetry, leading to the emergence of strong ASOC. Angle-resolved photoemission spectroscopy (ARPES) has confirmed the presence of a fully spin-polarized Dirac surface state, consistent with theoretical predictions \cite{Duan2019, Jeng2016, Bian2016, Hasan2016}. Scanning tunneling microscopy (STM) and quasiparticle interference (QPI) studies further indicate the existence of spin-polarized Dirac TSSs on the Pb-terminated surface, which could potentially become fully gapped at low temperatures \cite{Chuang2016, Multer2021, Lu2020}. However, with regard to the superconducting order parameter, there is currently no clear experimental evidence distinguishing the superconducting gap or/and critical temperature $T_c$ between the surface and bulk states. This contrasts with theoretical expectations, which predict that the superconducting properties of the surface and bulk should differ\cite{Su-Yang2014,Jin-peng2015,Dai2017}. One possible explanation for the lack of a clear distinction between the surface and bulk superconducting states is that the extent of inversion symmetry breaking in PbTaSe$_2$ under ambient pressure may not be sufficient to induce a significant ASOC, thus preventing a clear separation between the surface and bulk states.

Recently, high-pressure transport experiments on PbTaSe$_2$ have revealed two distinct types of superconductivity at high pressures, suggesting the presence of a first-order structural transition (Lifshitz transition) that may enhance the antisymmetric spin-orbit coupling (ASOC). As schematically illustrated in the main panel of Figure 1, at a critical pressure $P_c\simeq 2.5$ kbar, the $T_c$ experiences a stepwise drop or first-order-like transition\cite{Bud'ko2016, Murtaza2023, Hanasaki2022}. The superconducting $T_c-P$ phase diagram is divided into two regions: SC I and SC II \cite{Bud'ko2016}. In our previous report, it reveals the existence of two distinct $T_c$ in SC II region \cite{Ren2021}.  These findings provide an intriguing opportunity to explore ASOC-enhanced topological superconductivity in this material. In this work, we extend our previous research by systematically measuring the dependence of Andreev reflection spectra of PbTaSe$_2$ in SC II phase on magnetic field. Our goal is to search for signs of Majorana zero modes (MZMs) in ARS and to explore the physical characteristics of the superconducting order parameter in the SCII phase under the influence of ASOC strength.

\emph{Experimental Details:} High-quality single crystals of PbTaSe$_2$ were synthesized using the chemical vapor transport method, with iodine as the transport agent, a technique well-documented in previous literature\cite{Chen2016}. To probe the superconducting properties of PbTaSe$_2$, we utilized soft point contact Andreev reflection spectroscopy (SPCARs). This technique involved creating a soft point contact junction by attaching a 16$\mu$m diameter platinum wire, coated with silver paint (4929N DuPont), to the freshly cleaved surface of the PbTaSe$_2$ crystal under ambient conditions. This ensured a pristine contact surface, critical for accurate spectral analysis. The crystal was placed within a BeCu/NiCrAl pressure cell, using Daphne 7373 as the pressure medium to ensure a stable and uniform pressure environment, which is crucial for consistent experimental conditions.

At low-$T$s the values of hydrostatic pressure were determined by the superconducting critical temperatures $T_c$ of lead (Pb).  Andreev reflection conductance curves, $G(V)$, were recorded using a lock-in technique with a four-probe measurement, incorporating a Keysight B2901A current source for DC and an MFLI 500 kHz lock-in amplifier for AC currents. Comprehensive temperature- and field-dependent resistivity and spectroscopy measurements were conducted down to 0.3 K in an Oxford Instruments 3-He cryostat equipped with a 12 Telsa magnet.

\begin{figure}
\centering
\includegraphics[scale=0.16]{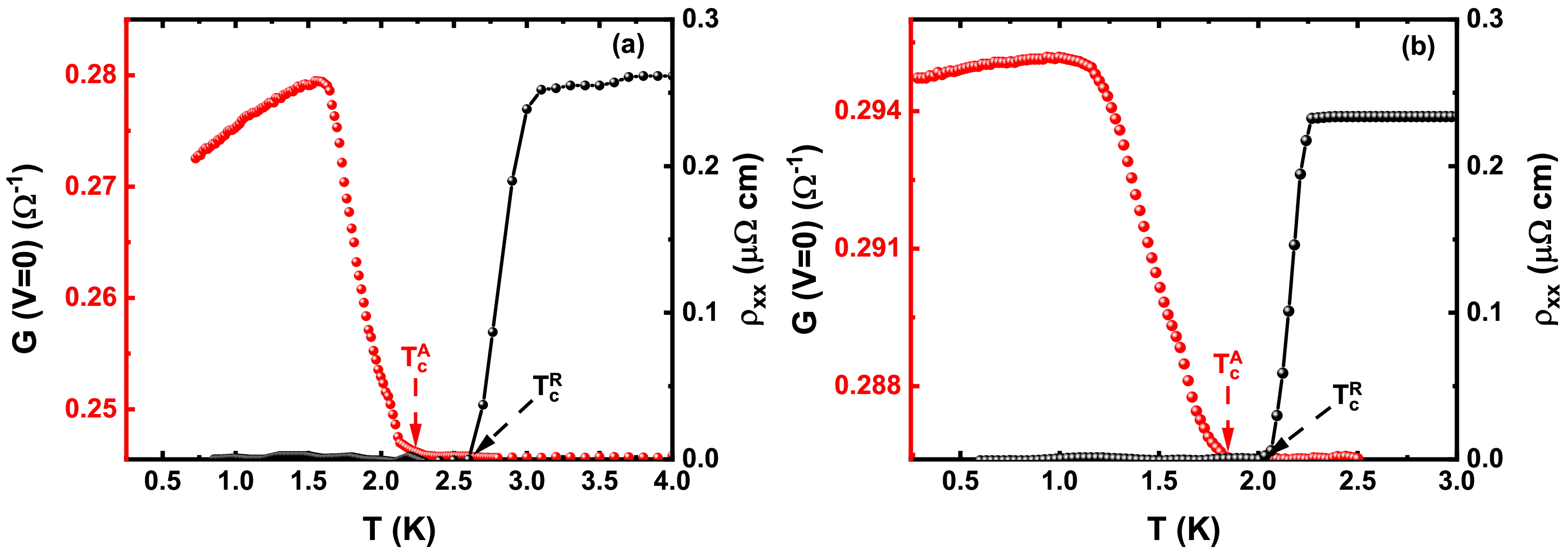}
\caption{The raw data of zero-bias Andreev conductance $G(V=0)$ and in-plane resistivity $\rho_{xx}$ as functions of temperature $T$ at pressures of 4.2 kbar (a) and 6.6 kbar (b). The arrows mark the Andreev temperature $T^A_c$ and zero-$\rho_{xx}$ temperature $T^R_c$, respectively. }
\end{figure}

\emph{Results and Discussions:} We performed a combined soft point-contact Andreev reflection (PCAR) spectra $G(0)$ with in-plane resistivity $\rho_{xx}$ measurements under various hydrostatic pressures $P$.  Figure 2 shows the zero-bias conductance $G(V=0)$ and $\rho_{xx}$ as functions of temperature $(T)$ near the superconducting transition region.  As shown in Fig. 2(a), at $P\simeq 4.2 $ kbar $ > P_c$, the Andreev temperature $T_c^A$, defined as the temperature at which the Andreev conductance features disappear and the conductance spectrum becomes indistinguishable from that of the normal state, the genuine superconducting gap opening/closing temperature, is about 2.20 K. For a comparison, the $T_c^R\simeq 2.55$ determined by zero-resistivity $\rho_{xx}$ measurement.  The same is true for $P\simeq 6.6 $ kbar, that $T_c^A\simeq 1.75$ K and $T_c^R\simeq 2.10$ K, yielding a $\Delta T_c\simeq 0.35$ K.  Here, we emphasize that the simultaneous PCAR and $R-T$ measurements in our experiment eliminated random errors in the pressure and $T$ determination. This is consistent with the observations in our previous study \cite{Ren2021}.  Noted, magnetization measurements by Kaluarachchi et al. using ac magnetic susceptibility also show a similar $\Delta T^c=0.40$ K difference between the critical temperature $T_c^M$ determined by magnetization and $T_c^R$ \cite{Bud'ko2016}.

Recent studies suggest the existence of two $T_c$ values corresponding to the bulk and new surface-like superconducting states, with $T_c^A$ potentially representing the surface superconducting state. As a non-centrosymmetric superconductor, PbTaSe$_2$ supports topological surface states that could harbor Majorana zero modes (MZMs). However, as shown in Fig 2, there is no any trace of a zero-bias conductance peak, suggesting the absence of MZMs in the SCII phase.  Increased pressure enhances the antisymmetric spin-orbit coupling (ASOC), which manifests as spin splitting and degeneracy lifting in bands near the Fermi surface. This, in turn, leads to the separation of surface states from bulk states.

A ZBCP in the conductance junction spectrum is widely regarded as a manifestation of a $p$-wave nodal gap in the superconducting order parameter \cite{GuQQ}. Figure 2 shows the absence of a zero-bias conductance peak (ZBCP) in the Andreev conductance at any temperature, suggesting the absence of an intrinsic spin-triplet $p$-wave component in the system under such pressures. The persistent presence of the zero-bias conductance dip in the high-pressure region is strong evidence against the dominance of a $p$-wave order parameter in the SC-II regime. Indeed, we measured $T$-dependent point contact spectra of PbTaSe$_2$ under $P$s, and the presentative PCS data under $P\simeq 4.2$ kbar are shown in Fig. 3(a).

To quantitatively assess the nature of the superconductivity in the surface state of PbTaSe$_2$ in the SC-II state, we determine the gap function $\Delta$ by fitting the PCAR spectra to a generalized two-dimensional Blonder-Tinkham-Klapwijk (BTK) model \cite{BTK,JapPRL}. Examples of the normalized $G(V)$ curves and their BTK fits at the selected $T$ values are shown in Fig.3(a) for $P\simeq 4.2$ kbar. The use of a single isotropic $s$-wave gap in the BTK model (colored lines) yields best fits to the data, which capture very well all the important features of the experimental $G(V)$ curves. The analyses yield a set of $T$-dependent fitting parameters, especially the gap size, $\Delta(T)$ , at different $T_s$.  Fig. 3(b) are the extracted gap values as functions of $T$. The resulting $\Delta(T)$ are well described by the BCS gap function: $\Delta(T) = \Delta_0 \tanh(\alpha \sqrt{T_c/T -1}$ with $\alpha\simeq 1.83$ is the coupling strength. Using this formula, we obtained the zero-$T$ gap $\Delta_0 = 0.50$ meV, and $T_c = 2.25$ K, fully consistent with $T^A_c$ determined from the zero-bias conductance $G(0)$.  The gap ratio $2\Delta_0/k_B T_c\simeq 5.16$ strongly suggest that the proximity-induced superconductivity in the surface states of PbTaSe$_2$ is in the strong-coupling regime.

\begin{figure}
\centering
\includegraphics[scale=0.16]{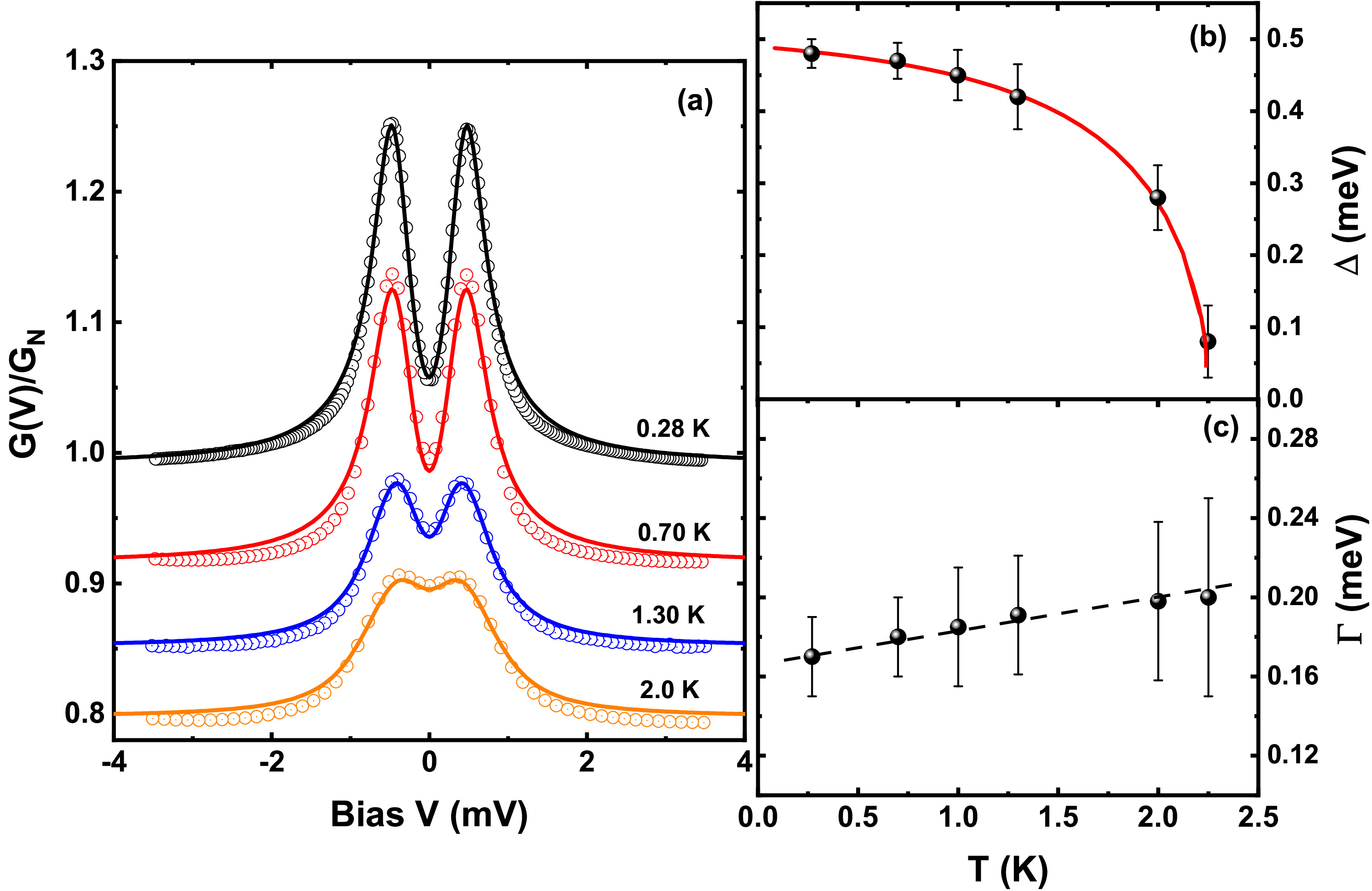}
\caption{ (a) The temperature evolution of the normalized differential conductance $G(V)/G_N$ under a hydrostatic pressure of 4.2 kbar. The corresponding colored solid curves are their BTK single $s$-wave-gap fits at various $T$s. The experimental data and their fitting curves are shifted downwards for clarity, except the top one.  (b) BTK-fitted superconducting gap $(\Delta)$ as a function of $T$. The solid red curve is the BCS $\Delta-T$ fit with $\Delta_0$ and $T_c$ as the fitting parameters.  (c) The extracted energy broadening factor $\Gamma$ (Dynes factor) as a function of $T$. The dashed line is a guide for the eye.}
\end{figure}

\begin{figure}
\centering
\includegraphics[scale=0.18]{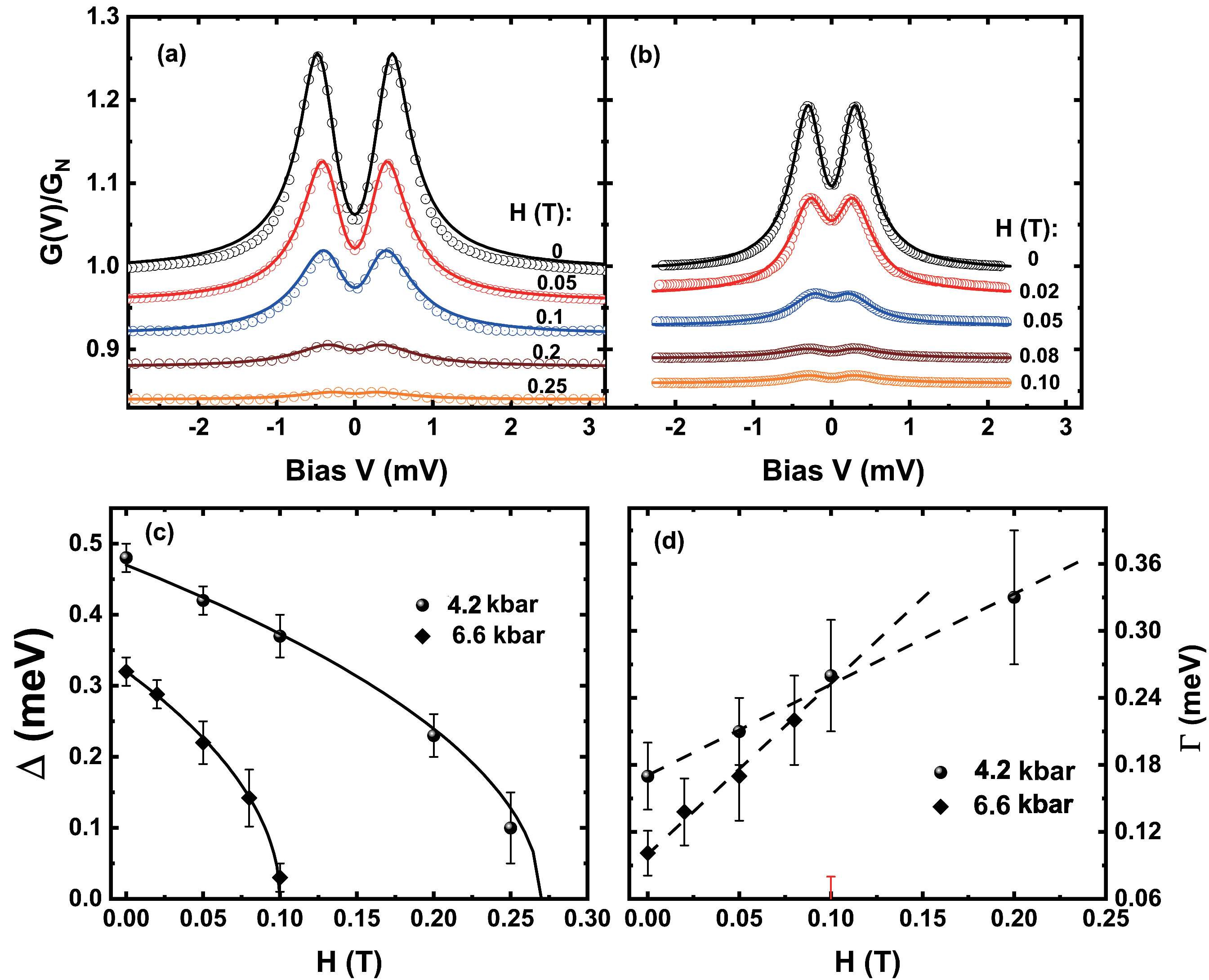}
\caption{Magnetic field $H$ dependence of the normalized differential conductance $G(V)/G_N$ at $T=0.28$ K under a hydrostatic pressure of 4.2 kbar (a) and 6.6 kbar (b).  The corresponding colored solid curves are their BTK single $s$-wave gap fits in various $H$s. (c) The resulted superconducting gap $\Delta$ as a function of $H$ under labeled pressures.  The solid black curves are the $\Delta-H$ fits with the formula: $\Delta(T,H)=\Delta(T,0) \sqrt{1-H/H_{c2}}$, where $H_{c2}$ is the upper critical magnetic field. (d) The extracted energy broadening factor $\Gamma$ as a function of $H$.  The dashed lines are the guide for the eye.}
\end{figure}

Theoretically, it was proposed to search MZM in vortex core on the isolated superconducting surface \cite{Tanaka2018}.  Recently, strong evidence of Majorana zero modes inside the magnetic field-induced vortices have been observed by scanning tunneling microscope/spectroscopy on the vortex core of $s_{\pm}$-wave iron-based superconductor \cite{Stevan2014,Wang2018,Zhang2018,Liu2020,Kong2021,Peng2021}.  Similarly, we examine the effect of magnetic field on the soft PCS of PbTaSe$_2$ under various $P$s by applying magnetic fields perpendicular to the $ab$ plane. Figure 4 (a)-(b) show the measured Andreev reflection spectra of PbTaSe$_2$/Ag junction in various magnetic fields perpendicular to the $ab$-plane at the lowest $T=0.28$ K under a pressure of $P\simeq 4.2$ and 6.6 kbar, respectively.  As shown in Fig. 4(a), while under zero field, the measured Andreev conductance $G(V)/G_N$ shows a large double peak-deep dip at zero bias, it means the dominance of tunneling component in such Andreev conductance.  As applying $H=0.05$ T, there is no any trace of zero-bias conductance peak.  The same is true at $P=6.6$ kbar [Fig. 4(b)] that no any trace of ZBCP.  Magnetic field-induced zero-bias conductance peak as Majorana zero-energy mode in vortex core.

As increasing $H$, the intensity of the Andreev conductance spectra are suppressed toward that of the normal state under a $H\simeq 0.25$ Telsa.  The SPCS conductance curves at different fields still can be well fitted by this single $s$-wave BTK model as shown in Fig. 4(a) and the parameters $Z$ change little with $H$.  The resulted superconducting gap $\Delta$ as a function of $H$ at $T=0.28$ K are plotted in Fig. 4(c).  As shown, the gap value follow the theoretical prediction for a one-gap type-II superconductor in a magnetic field with the relation $\Delta(H) = \Delta_0 \sqrt{1 - H/H_{c2}}$ \cite{Lu2019}.  For example, at $T=0.27$ K, $\Delta(H=0)\simeq 0.49$ meV and $H_{c2}(\Delta=0)\simeq 0.27$ Telsa. It is noted that this $\Delta-H$ relationship reflect a three-dimension (3D) superconducting state, as predicted.  Pauli paramagnetic limitation $H_p=\Delta/\sqrt{2}\mu_B$ with $\mu_B$ the magnetic moment of electron. Given $\Delta\simeq 0.49$ meV under $P\simeq 3.1$ kbar, $H_p\simeq 5.2$ Tesla, far beyond the upper critical magnetic field $H_{c2}\simeq 0.26$ Tesla obtained.  This indicates the absence of spin-triplet component in such proximity-induced superconducting surface state.

\begin{figure}
\centering
\includegraphics[scale=0.32]{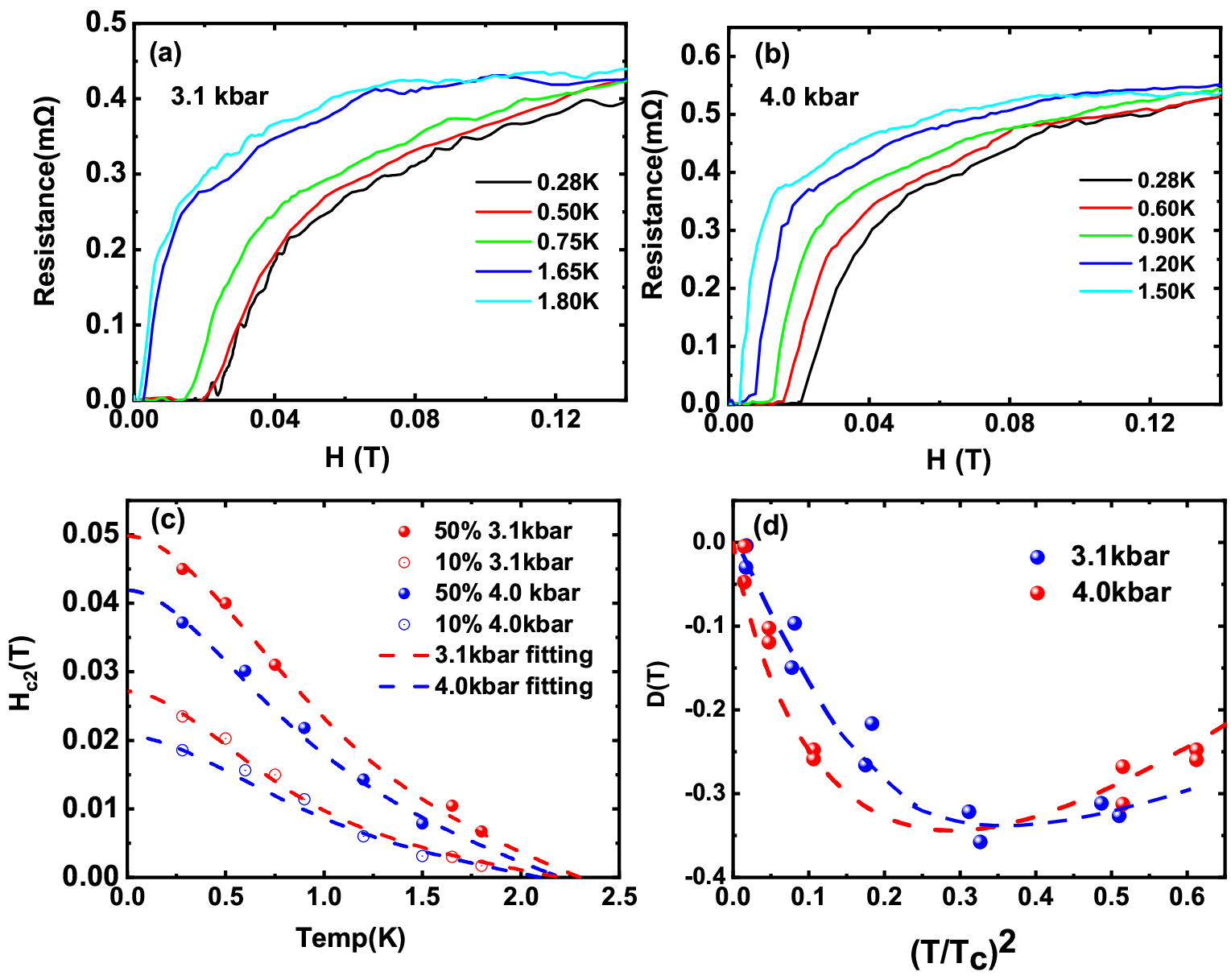}
\caption{Temperature dependence of the in-plane resistance $R_{xx}$ as a function of magnetic field $H$ parallel to $c$-axis under a pressure of 3.1 kbar (a) and 4.0 kbar. (c) The criterion-based  upper critical field $H_{c2}$ as a function of $T$. The dashed color curves are the two-band model-fitting ones. (d) The defined dimensionless parameter $D(T)$ as a function of reduced temperature $(T/T_c)^2$. The dashed color curves are a guide to the eye, demonstrating the trend of $D(T)$ with reduced temperature based on the $H_{c2}$ determination criterion.}
\end{figure}

To emphasize that the surface-like state probed by our SPCAR spectra represents a distinct superconducting state, different from the bulk superconducting state, we measured the magnetoresistance for $H//c$-axis under $P>P_c$.  As shown in Figure 5 (a) and (b), we used two criteria to determine the critical temperature during the superconducting transition: $T_c^{50\%}$ (the 10\% of the normal-state resistance) and $T_c^{offset}$ (the zero-resistance point). The data were fitted using a two-gap model \cite{Brandt1976,Gurevich,ShiZX}. Based on the extracted $H_{c2}(0)$ and $T_c(0)$, a dimensionless function is defined:
\begin{equation}
D(T)\equiv{\frac{H_c(T)}{H_c(0)}}-[1-(\frac{T}{T_c})^{2}],
\end{equation}
which reflects the superconducting pairing strength. In weakly coupled BCS superconductors like aluminum and tin, $D(T)$ is typically negative. As shown in Figure 5 (d), the curves derived from both the zero-resistance point and the 50$\%$ resistance transition point indicate that $D(T)$ for PbTaSe$_2$ in the SC-II phase is also negative, suggesting that the material behaves similarly to a weakly coupled superconductor. However, this observation contrasts with our Andreev reflection measurements, which show stronger coupling with $2\Delta / T_c k_B > 3.52 $. This discrepancy highlights that in the SCII phase, the bulk and surface-like superconductivity in PbTaSe$_2$ are independent states, each exhibiting different pairing strengths. Furthermore, point-contact measurements under applied pressure reveal a significant transition from weak to strong coupling, with the surface-like superconducting gap exceeding the bulk superconducting gap. Notably, the pressure at which this transition occurs coincides with the pressure at which the structural phase transition happens. This provides further support for the conclusion that when pressure exceeds 2.5 kbar, PbTaSe$_2$ hosts two decoupled superconducting states.
\emph{Conclusions:} We have performed combining point-contact Andreev reflection spectroscopy with bulk resistivity measurements under quasi-hydrostatic pressures to investigate the effect of ASOC on the superconductivity of PbTaSe$_2$. With increasing pressure, PbTaSe$_2$ experiences a structural Lifshitz transition with an enhanced ASOC in the high-$P$ phase. In the high-$P$ region, the superconducting-gap opening temperature $T^A_c$ of the surface state is found to be consistently lower than the bulk resistive transition temperature $T^R_c$. The observation signals a surface-bulk separation and a proximity-induced superconducting TSS in PbTaSe$_2$. The induced superconductivity in the PbTaSe$_2$ TSSs is BCS-like in strong-coupling limit.  The absence of Andreev bound state hosting Majorana fermions imposes a constraint on the proposal for realizing Majorana bound states in noncentrosymmetric superconductor.

\textbf{Acknowledgements:} This work was supported by the National Natural Science Foundation of China.  G. F. Chen thanks the support of the Ministry of Science and Technology of China (Grant No. 2022YFA 1403903). Y. Zhao thanks financial support from ``16th Graduate Research Innovation Project'' of Yunnan Unvercity.

\end{document}